\documentstyle[aps,preprint]{revtex}
\begin{document}
\draft
\title{ Testing the meson cloud in the nucleon
in Drell-Yan processes}

\author {A. Szczurek}
\address{Institute of Nuclear Physics, PL-31-342 Cracow, Poland}

\author{M. Ericson}
\address{ Institut de Physique Nucleaire et IN2P3, CNRS\\
Universite Claude Bernard,
Lyon, F-69622 Villeurbanne cedex, France\\
and Theory Division, CERN, CH-1211 Geneva 23, Switzerland}

\author{H. Holtmann and J. Speth}
\address{Institut f\"ur Kernphysik, Forschungszentrum J\"ulich,
D-52425 J\"ulich, Germany}

\date{\today}
\maketitle
\begin{abstract}
We discuss the present status of the $\overline{u}-\overline{d}$
asymmetry in the nucleon and analize the quantities which are best
suited to verify the asymmetry.
We find that the Drell-Yan asymmetry is the quantity insensitive
to the valence quark distributions and very sensitive to the flavour
asymmetry of the sea.
We compare the prediction of the meson cloud model with different
experimental data including the Fermilab E772 data and recent data of
the NA51 Collaboration at CERN and make predictions for the planned
Drell-Yan experiments.
\end{abstract}


\section{Introduction}

The deviation of the Gottfried Sum Rule from its classical value
\cite{G67}

\begin{equation}
S_{G} = \int_{0}^{1} [ F_{2}^{p}(x) - F_{2}^{n}(x) ] \frac{dx}{x}
            = \frac{1}{3}
\label{GSR}
\end{equation}
observed by the New Muon Collaboration (NMC) at CERN \cite{A91,A'94}
has created a large interest on the possible sources of the
violation.  In the NMC experiment the neutron structure function
which enters the sum rule is deduced from deep inelastic scattering
off deuterium. It could be biased by nuclear two-body effects which
were ignored in the NMC analysis.  While shadowing effects
\cite{BK92,Z92A,KH93,MT93} cause the real value of the Gottfried Sum
Rule to be even smaller than the value given by NMC, the
anti-shadowing effect, due to the presence of virtual mesons which
bound the deuteron, tends to restore the classical value
\cite{KU91,MT93}.  A recent unfolding of the shadowing and
anti-shadowing effects \cite{MT93} suggest, however, a genuine
violation. This can be understood as consequence of an internal
asymmetry $\overline{d} > \overline{u}$ of the quarks in the proton
(the opposite asymmetry is expected for the neutron if the
proton-neutron charge symmetry holds). The asymmetry has been
confirmed recently by the NA51 collaboration group at CERN
\cite{B94}.  Since at large $Q^2$ the perturbative QCD evolution is
flavour independent and, to leading order in $logQ^2$, generates
equal number of $\overline{u} u$ and $\overline{d} d$ sea quarks
nonperturbative effects must play an important role here. The effect
of the asymmetry has been predicted by the models in which the
physical nucleon contains an admixture of the $\pi N$ and $\pi
\Delta$ components in the Fock expansion \cite{SSG94}. The predicted
effect of the asymmetry agrees\cite{HNSS94} with that deduced from
the NA51 experiment.

The idea of the $\overline{u}-\overline{d}$ asymmetry is not new.
It was considered a decade ago by Ito et al.\cite{IFJ81}
as a possible explanation for the slope of the rapidity distribution
of dilepton production in the proton-Pt
collision at Fermilab. An alternative interpretation
stimulated by the discovery of the EMC
effect invoked the enhancement of the nuclear sea in Pt with respect
to a collection of free nucleons \cite{ET84}.
In general both effects can coexist.

In the advent of high precision data on the dilepton production in
the proton-proton and proton-deuteron scattering \cite{G92}, it is
interesting to review the present status of our knowlegde on the
$\overline{u}-\overline{d}$ asymmetry.
We will discuss the quantities which should most unambiguously
confirm the asymmetry and allow for verification of different
theoretical concepts.
We also confront the prediction of the meson cloud model with
the existing data.

\pagebreak
\section{Gottfried Sum Rule}
\label{secGSR}

\mbox{ }\indent The Gottfried Sum Rule\cite{G67} (GSR) addresses the
value of the integral over $x$ of the difference of the $F_{2}(x)$
structure function of the proton ($p$) and neutron ($n$).  It is
written\footnote{The structure functions $F_{2}(x)$ and the quark
distribution functions $q(x)$ are, of course, functions of
$Q^{2}$. However, the $Q^{2}$ dependence is suppressed to keep the
expressions from being too cumbersome.} as
\begin{eqnarray}
\int_{0}^{1}\left[F_{2}^{p}(x)-F_{2}^{n}(x)\right] \;\
\frac{dx}{x} \;\ &=&
\int_{0}^{1}
\frac{4}{9}\left[u_{p}^{v}(x)-u_{n}^{v}(x)\right] +
\frac{1}{9}\left[d_{p}^{v}(x)-d_{n}^{v}(x)\right]
\\ \nonumber
&+&
2 \left\{
\frac{4}{9} \left[\overline{u}_{p}(x)-\overline{u}_{n}(x)\right] -
\frac{1}{9} \left[\overline{d}_{p}(x)-\overline{d}_{n}(x)\right]
\right\} \; dx \; ,
\end{eqnarray}
where $u_{p}^{v}(x)\equiv u_{p}(x)-\overline{u}_{p}(x)$, etc.
Baryon number conservation reduces the expression further to
\begin{equation}
S_{G} = \frac{1}{3} +
\int_{0}^{1} \;
\frac{8}{9} \left[\overline{u}_{p}(x)-\overline{u}_{n}(x)\right] -
\frac{2}{9} \left[\overline{d}_{p}(x)-\overline{d}_{n}(x)\right]
\; dx \; .
\label{GSR_viol1}
\end{equation}
As seen from Eq.(\ref{GSR_viol1}) the valence quarks do not
influence the Gottfried Sum Rule violation although they are of
crucial importance for the GSR integrand.  Assuming further charge
symmetry for the nucleon sea, i.e.,
$\overline{u}_{p}(x)=\overline{d}_{n}(x)$, etc. and making the
customary assumption that $\overline{u}_{p}(x)=\overline{d}_{p}(x)$,
one finds the classical value of $1/3$.  The NMC experiment
\cite{A91} of the relevant structure functions, over the interval
$0.004\le x\le 0.8$, yielded when extrapolated to $0\le x\le 1$,
\begin{equation}
\int_{0}^{1}\left[F_{2}^{p}(x)-F_{2}^{n}(n)\right] \;\
\frac{dx}{x} = 0.24\pm 0.016 \; ,
\label{exp1}
\end{equation}
at $Q^{2}=5$~GeV$^{2}$.
It should be noted that QCD corrections do not play any role here.
While the leading order corrections to the Gottfried Sum Rule cancel,
the higher order corrections are negligibly small.

The integrand of the Gottfried Sum Rule has been obtained from the
ratio $R = F_{2}^{n}/F_{2}^{p}$ and the deuteron $F_{2}^{d}$
structure function
\begin{equation}
F_{2}^{p}(x) - F_{2}^{n}(x) = 2 F_{2}^{d}(x)
\frac{1 - R(x)}{1 + R(x)}  \; .
\end{equation}
In the first evaluation of GSR the deuteron structure function was
taken from a global fit to the results of earlier experiments.
In the meantime NMC has published values of $F_{2}^{p}$ and
$F_{2}^{d}$ of its own accurate measurements at low $x$ \cite{A92}.
The old value of GSR \cite{A91} has been updated \cite{A'94} to
\begin{equation}
\int_{0}^{1}\left[F_{2}^{p}(x)-F_{2}^{x}(n)\right] \;\
\frac{dx}{x} = 0.235\pm 0.026 \; .
\label{exp2}
\end{equation}
The total error here is larger than in (\ref{exp1}) due to a more
extensive examination of the systematic uncertainties.
The quoted error bar does not include the effects of
shadowing and antishadowing.

In the most general case not only the so-called $SU(2)_{Q}$
\cite{F93} charge-symmetry is violated but also the isospin symmetry
between proton and neutron $SU(2)_{I}$. In this case the antiquark
distributions in proton and neutron can be expressed as:
\begin{eqnarray}
\overline{u}_{p}(x) &=&
\overline{q}(x)-\frac{1}{2}\Delta_{Q}(x)-\frac{1}{2}\Delta_{I}(x)
\; , \nonumber \\
\overline{d}_{p}(x) &=&
\overline{q}(x)+\frac{1}{2}\Delta_{Q}(x)-\frac{1}{2}\Delta_{I}(x)
\; , \nonumber \\
\overline{u}_{n}(x) &=&
\overline{q}(x)+\frac{1}{2}\Delta_{Q}(x)+\frac{1}{2}\Delta_{I}(x)
\; , \nonumber \\
\overline{d}_{n}(x) &=&
\overline{q}(x)-\frac{1}{2}\Delta_{Q}(x)+\frac{1}{2}\Delta_{I}(x)
\label{SU(2)_viol} \; .
\end{eqnarray}
The signs in front of $\Delta_{Q}(x)$ and $\Delta_{I}(x)$ have
been chosen to assure positivity of $\Delta_{Q}$ and $\Delta_{I}$ in
the case the asymmetries give the effect required by the NMC result.
With the parameterization (\ref{SU(2)_viol}) the Gottfried Sum Rule
(\ref{GSR_viol1}) can be written as
\begin{equation}
S_{G} =
\frac{1}{3} - \frac{2}{3} \Delta_{Q} - \frac{10}{9} \Delta_{I}
\; .
\label{GSR_viol2}
\end{equation}
The factor 10/9 shows the sensitivity to the $SU(2)_{I}$ symmetry
violation.  The violation of the Gottfried Sum Rule exclusively via
breaking the $SU(2)_{Q}$ symmetry requires $\Delta_{Q} = 0.14 \pm
0.03$.  On the other hand violation of GSR exclusively via breaking
the $SU(2)_{I}$ symmetry requires $\Delta_{I} = 0.08 \pm 0.02$.

It has been argued in Ref.\cite{MSG93} that both the effects of
asymmetric $SU(2)$ sea and the proton-neutron isospin symmetry
breaking will be very difficult to disentangle as they may, in
general, lead to very much the same behaviour both in deep inelastic
scattering and Drell-Yan processes. A careful analysis\cite{F93}
based on the $\sigma$-term and scale Ward identity suggests rather
$\Delta_{Q} \gg \Delta_{I}$.  Therefore in the further analysis we
will neglect the effect of the $SU_{I}(2)$ symmetry violation.

Since the standard Altarelli-Parisi \cite{AP77} evolution equations
generate equal number of $\overline{u}u$ and $\overline{d}d$ pairs,
one does not expect strong scale dependence of the GSR. The two-loop
evolution gives a rather negligible effect \cite{RS79}.  The Pauli
exclusion principle leads to some interference phenomena which
produces only a small asymmetry \cite{RS79}.

The answer likely lies with more complicated nonperturbative physics.
The presence of the pion cloud in the nucleon gives a natural
explanation of the excess of $\overline{d}$ over $\overline{u}$.  It
has been extensively analyzed in a series of papers
\cite{HM90,SST91,MTS91,K191,HSB91,Z92,SS93} with the result for the
GSR being dependent on the details of the model, especially on vertex
form factors. Restricting the form factors by fitting to the cross
sections for high-energy production of neutrons and $\Delta^{++}$
yields the GSR \cite{Z92,SSG94,HSS94} in rough agreement with that
obtained by NMC \cite{A91,A'94}.

The flavour content of the sea in the nucleons can be tested in
Drell-Yan experiments, which is the subject of the present paper.
In the first phenomenological analyses the antiquark distributions
were typically defined as
\begin{equation}
\overline{u}(x) = \overline{q}(x) - \frac{1}{2} \Delta(x) \qquad
{\rm and}\qquad
\overline{d}(x) = \overline{q}(x) + \frac{1}{2} \Delta(x) \; .
\label{asymdecomp}
\end{equation}
Different assumptions on $\Delta (x)$ lead to different predictions
for the Drell-Yan production rate.  For instance, the initial
parameterization \cite{ES91} used $\Delta(x)=A(1-x)^{k}$, which
placed the $\overline{d}-\overline{u}$ difference at large $x$
($x>0.05$), led to very large values for
$\overline{d}_{p}(x)/\overline{u}_{p}(x)$ for $x\ge 0.1$.  These
large ratios have been ruled out by a recent reanalysis of earlier
Drell-Yan data \cite{GMA92}.  More recent\cite{MSR94}
parameterizations have a form $\Delta(x)=A_\Delta
x^\eta(1-x)^{\eta_S}(1+\gamma x)$, which places the bulk of the
difference at smaller $x$.

It is expected that future QCD lattice gauge calculations will be
able to generate the quark structure of the nucleon and evaluate the
$\overline{d}_{p}(x)$ and $\overline{u}_{p}(x)$ distributions.  At
present one has to rely on phenomenological models consistent with
our knowledge in other branches of hadronic physics.  The meson
cloud model \cite{HSS93,HSS94} seems to satisfy this
criterion. Furthermore it is worth mentioning in this context that
importance of pion loop effects for nucleon properties has been
demonstrated in a recent lattice QCD calculation \cite{CL93}.

\pagebreak
\section{Meson cloud model of the nucleon}
\label{MCM}

In this section we briefly sketch the meson cloud model (MCM) of the
nucleon\cite{SS93,HSS93,HSS94} and present its prediction for the
asymmetry of the light sea antiquarks.  In this model the nucleon is
viewed as a quark core, termed a bare nucleon, surrounded by the
mesonic cloud. The nucleon wave function can be schematically
written as a superposition of a few principle Fock components (only
$\pi N$ and $\pi\Delta$ are shown explicitly)
\begin{eqnarray}
&&|p\rangle_{phys}=\sqrt{Z}\Big[ \; |p\rangle_{core} \nonumber\\
&&+
    \int dy\,d^2\vec k_\perp\phi_{N\pi}(y,\vec k_\perp)\Big(
        \sqrt{1\over 3} |p\pi^0;y,\vec k_\perp\rangle
      + \sqrt{2\over 3} |n\pi^+;y,\vec k_\perp\rangle\Big) \nonumber\\
&&+
    \int dy\,d^2\vec k_\perp\phi_{\Delta\pi}(y,\vec k_\perp)\Big(
        \sqrt{1\over 2} |\Delta^{++}\pi^-;y,\vec k_\perp\rangle
      - \sqrt{1\over 3} |\Delta^{+}\pi^0;y,\vec k_\perp\rangle
      + \sqrt{1\over 6} |\Delta^{0}\pi^+;y,\vec k_\perp\rangle\Big)
    \nonumber\\
&&+  \ldots \;\ \Big] .
\label{Fock}
\end{eqnarray}
with $Z$ being the wave function renormalization constant which can
be calculated by imposing the normalization condition $\langle
p|p\rangle=1$. The $\phi(y,\vec k_\perp)'s$ are the light cone wave
functions of the $\pi N$, $\pi\Delta$, etc. Fock states, $y$ is the
longitudinal momentum fraction of the $\pi$ (meson) and $\vec
k_\perp$ its transverse momentum.

It can be expected that the structure of the bare nucleon (core) is
rather simple.
Presumably, it can be described as a three quark system in the static
limit. Of course, in the deep inelastic regime at higher $Q^2$
additional sea of perturbative nature is created unavoidably by
the standard QCD evolution.

The model includes all the mesons and baryons required in the
description of the low energy nucleon-nucleon and hyperon-nucleon
scattering, i.e. the $\pi$, $K$, $\rho$, $\omega$, $K^*$ and the $N$,
$\Lambda$, $\Sigma$, $\Delta$ and $\Sigma^*$.
In contrast to other approaches in the literature the
model ensures charge conservation, baryon number and momentum sum rules
\cite{HSS93} by construction.

The main ingredients of the model are the vertex coupling constants,
the parton distribution functions for the virtual mesons and baryons
and the vertex form factors which account for the extended nature
of the hadrons. The coupling constants are assumed to be related via
$SU(3)$ symmetry which seems to be well established from low-energy
hyperon-nucleon scattering.

It was suggested in Ref.\cite{Z92} to
use the light cone meson-baryon vertex form factor
\begin{equation}
F(y,k_{\perp}^2) =
\exp\left[-\frac
{M_{MB}^2(y,k_{\perp}^2) - m_N^2}{2 \Lambda_{MB}^{2}}\right],
\end{equation}
where $k_{\perp}$ is the transverse momentum of the meson and
$M_{MB}(y,k_{\perp}^2)$ is the invariant mass of the intermediate
two-body meson-baryon Fock state,
\begin{equation}
M_{MB}^2(y,k_{\perp}^2) =
\frac{m_B^2 + k_{\perp}^2} {1-y} \; + \;
\frac{m_M^2 + k_{\perp}^2} {y} \; .
\end{equation}
By construction, such form factors assure the momentum sum rule
\cite{Z92,HSS93}. The parameters $\Lambda_{MB}$ are the principal
nonperturbative parameters of the model. They have been determined
from an analysis of the $p\rightarrow n,\Delta,\Lambda$ fragmentation
spectra \cite{HSS94} using light
cone flux functions \cite{HSS94} ($\Lambda_{\pi N}^{2}=$ 1.08
$GeV^{2}$ and $\Lambda_{\pi \Delta}^{2}=$ 0.98 $GeV^{2}$).
With these parameters
the pion exchange model gives a very good description of the spectra.

In practice the probability of the Fock components with
strange particles is rather small. For instance the probability
to find a $K\Lambda$ state in the nucleon is about 1\%,
whereas the probability to find a $\pi N$ state is $0.18$.
In all applications in the present paper the higher Fock states
involving strange particles can be neglected.

The $x$ dependence of the structure functions in the meson cloud
model can be written as a sum of components corresponding to the
expansion given by Eq.(\ref{Fock}).
\begin{equation}
F_{2}^{N}(x) = Z \left[ F_{2,core}^{N}(x) +
 \sum_{MB} \left( \delta^{(M)} F_{2}(x) +
                  \delta^{(B)} F_{2}(x) \right) \right] .
\label{dyratio}
\end{equation}
The contributions from the virtual mesons can be written as a
convolution of the meson (baryon) structure functions and its
longitudinal momentum distribution in the nucleon \cite{S72}
\begin{equation}
\delta^{(M)} F_{2}(x)
  = \int_{x}^{1} dy f_{M}(y) F_{2}^{M}({x\over y}) .
\label{mesons1}
\end{equation}
The same is true for the virtual baryons
\begin{equation}
\delta^{(B)} F_{2}(x)
  = \int_{x}^{1} dy f_{B}(y) F_{2,core}^{B}({x\over y}) .
\label{mesons}
\end{equation}
In a natural way $f_{M}(y)$ and $f_{B}(y)$ are related via
\cite{HSS93}
\begin{equation}
f_{B}(y) = f_{M}(1-y) \; .
\end{equation}
Eq. (\ref{mesons1}) (and also Eq. \ref{mesons}) can be written in an
equivalent form in terms of the quark distribution functions
\begin{equation}
\delta^{(M)} q_{f}(x)
= \int_{x}^{1} f_{M}(y) q_{f}^{M}({x\over y}){dy\over y} .
\end{equation}
The longitudinal momentum distributions (splitting functions, flux
factors) of virtual mesons (or baryons) can be calculated assuming a
model of the vertex and depend on the coupling constants and vertex
form factors. Further details can be found in
Refs.\cite{HSS93,HSS94}.

The parton distributions "measured" in pion-nucleus Drell-Yan
processes \cite{SMRS92} are used for the mesons.  The deep-inelastic
structure functions of the bare baryons, $F_{2,core}^{N}(x,Q^2)$,
$F_{2,core}^{B}(x,Q^2)$ are in principle unknown. In practical
calculations it is usually assumed $F_{2,core}^{N}(x,Q^2) =
F_{2,phys}^{N}(x,Q^2)$ \cite{MTS91,HS92}, which is not fully
consistent. Recently \cite{HSS95Q}, we have extracted
$F_{2,core}^{N}$ by fitting the quark distributions in the bare
nucleon, together with (parameter-free) mesonic corrections, to the
experimental data on deep-inelastic scattering:
\begin{itemize}
\item [(a)] $F_{2}^{p}(x,Q^2) - F_{2}^{n}(x,Q^2)$ \cite{A'94},
\item [(b)] $F_{2}^{n}(x,Q^2) / F_{2}^{p}(x,Q^2)$ \cite{A'94},
\item [(c)] $F_{3}^{\nu N}(x,Q^2)$ \cite{Le93},
\item [(d)] $\overline{q}(x,Q^2)$ \cite{M92}.
\end{itemize}
The following simple parameterization has been used for the quark
distributions in the bare proton at the initial scale $Q_{0}^{2}$ =
4 (GeV/c)$^2$. Note, that we have used $\bar u$-$\bar d$ symmetric
sea quark distribution for the core.
\begin{eqnarray}
x u_{v,core}(x,Q_{0}^{2}) &=&
N_{u} x^{0.38} (1-x)^{2.49} (1+10.5 x) \; ,
\\ \nonumber
x d_{v,core}(x,Q_{0}^{2}) &=&
N_{d} x^{0.07} (1-x)^{4.63} (1+150 x) \; ,
\\ \nonumber
x S_{core}(x,Q_{0}^{2}) &=& 0.17(1-x)^{13.8} \; ,
\label{baredis}
\end{eqnarray}
where
\begin{equation}
S_{core} =
u_{s,core} = \overline{u}_{s,core} =
d_{s,core} = \overline{d}_{s,core} =
2 s_{s,core} = 2 \overline{s}_{s,core} .
\end{equation}
The details of the fit and a comprehensive discussion of DIS will be
given in Ref.\cite{HSS95Q}.  An example of the fit to $F_{2}^{p}(x)
- F_{2}^{n}(x)$ is shown in Fig.1.
The resulting total sea quark distribution is compared in Fig.2
with an experimental sea quark distribution obtained from (anti)neutrino
induced reactions \cite{M92}.
The so-extracted parameterization of the quark distributions
in the bare baryons can be used to calculate the cross-sections for
both the lepton deep-inelastic scattering and for the Drell-Yan
processes. In the present article
we will present predictions(!)  for the Drell-Yan dilepton
production in elementary nucleon-nucleon collisions as well as for
the nucleon-nucleus collisions.

\pagebreak
\section{Drell-Yan processes}
\label{DrellYan}

\mbox{ }\indent The Drell-Yan process\cite{DY71} involves the
electromagnetic annihilation of a quark (antiquark) from the
incident hadron $A$ with an antiquark (quark) in the target hadron
$B$.  The resultant virtual photon materializes as a dilepton pair
($\ell^{+}\ell^{-}$) with muons being the pair most readily detected
in experiments.

Ellis and Stirling have shown that the measurement of Drell-Yan
cross sections in the proton-proton and proton-deuteron collisions
provides information on the
$\overline{d}_{p}(x)/\overline{u}_{p}(x)$ ratio \cite{ES91}.

The cross section for the DY process can be written as
\begin{equation}
\frac{d\sigma^{AB}}{dx_{1}dx_{2}} \;\ = \;\
\frac{4\pi\alpha^{2}}{ 9sx_{1}x_{2} }
K(x_{1},x_{2})\sum_{f}e_{f}^{2}
\left[q_{A}^{f}(x_{1})\overline{q}_{B}^{f}(x_{2})+
     \overline{q}_{A}^{f}(x_{1})q_{B}^{f}(x_{2})\right] ,
\label{DY}
\end{equation}
where $s$ is the square of the center-of-mass energy and $x_{1}$ and
$x_{2}$ are the longitudinal momentum fractions carried by the
quarks of flavour $f$.  The $q_{A}^{f}(x_{1})$ ($\bar
q_{A}^{f}(x_{1})$) and $q_{B}^{f}(x_{2})$ ($\bar q_{B}^{f}(x_{2})$)
are the (anti-)quark distribution functions of the beam and target,
respectively.  The factor $K(x_{1},x_{2})$ accounts for the
higher-order QCD corrections that enter the process.  Its value over
the kinematic range where experiments are carried out is typically
1.5.  The values of $x_{1}$ and $x_{2}$ are extracted from
experiment via
\begin{equation}
M^{2}=sx_{1}x_{2}\approx
  2P_{\ell^{+}}P_{\ell^{-}}(1-\cos\theta_{\ell^{+}\ell^{-}}) ,
\end{equation}
where $M$ is the mass of the dilepton pair, $P_{\ell^{+}}$ and
$P_{\ell^{-}}$ are the laboratory momenta of the leptons, and
$\theta_{\ell^{+}\ell^{-}}$ is the angle between their momenta
vectors.  The total longitudinal momentum of the lepton pair
(${P}_{\ell^{+}}+{P} _{\ell^{-}})_{L}$ fixes $x_{1}-x_{2}$ via
\begin{equation}
x_{1}-x_{2}\equiv x_{F}=
\frac{2\left(P_{\ell^{+}} + P_{\ell^{-}}\right)_{L}} {s} - 1 .
\end{equation}
In order to avoid spurious contributions to the DY yield from vector
meson decays, all measurements are made for $M>4$~$GeV$, and the
region $9\le M\le 11$~$GeV$ is excluded to avoid the $\Upsilon$
resonances.

The absolute value of the Drell-Yan cross section is biased by the
uncertainty in extrapolating from time-like to space-like values of
$Q^2$ when relating the Drell-Yan with deep-inelastic scattering
which involves the factor $K$ (see Eq.\ref{DY}).  In order to avoid
the uncertainty it is desirable to consider ratios
\cite{ET84,ES91,SSG94} rather than the absolute cross
sections. Whether the $K$-factors for the $pp$ and $pn$ collisions
are identical can be checked by calculating higher-order QCD
corrections.  In Fig.3 we show the $K$-factors calculated according
to the formalism developed in Ref.\cite{SMRS92} for two different
leading order quark distributions \cite{O91,MSR93}. Although the
$K$-factor depends on the input quark distribution as well as on
$x_1$ and $x_2$ its value is practically identical for proton-proton
and proton-neutron collisions.  The approximate equality $K_{pp} =
K_{pn}$ allows us to neglect the higher order corrections and limit
ourselves to the much simpler leading order analysis.

The quark distributions found from the procedure described in the
former section can be further tested by comparison with the
Drell-Yan E772 data \cite{GMA92} for the differential cross section
$M^{3} {d^2 \sigma}/{dx_{F} dM}$ for the dilepton production in the
$p+d$ collision. By fitting the $K$-factors the result of the
calculation can be compared with the experimental data. In order to
check the sensitivity to the $\bar u-\bar d$ asymmetry we have also
performed the calculation with symmetrized sea:
\begin{equation}
u_{s}(x) = \overline{u}_{s}(x) = d_{s}(x) = \overline{d}_{s}(x) =
\frac{\overline{u}(x) + \overline{d}(x)}{2} \; .
\label{symmetrization}
\end{equation}
In Fig.4 we show the result of the fit for our original model (solid
line, $\chi^2/N=1.15$) and the results obtained with the symmetrized
(Eq. \ref{symmetrization}) sea distribution (dashed line,
$\chi^2/N=1.42$). Altough there is some sensitivity to the $\bar
d-\bar u$ asymmetry, it can be easily compensated by a slightly
different normalization factor. For comparision we also show the
result obtained with the $MSR(S'_0)$ quark distributions
(dash-dotted line, $\chi^2/N=1.84$).

The present experimental data for the Drell-Yan processes in the
elementary nucleon-nucleon collisions suffer from rather low
statistics. Therefore at present one is forced to compare a
theoretical calculation with the proton-nucleus experimental data.
In the first approximation the cross section for the production of
the dilepton pairs in the proton-nucleus scattering can be expressed
in terms of the elementary $pp$ and $pn$ processes as
\begin{equation}
\sigma_{pA}^{DY} = Z\sigma_{pp}^{DY} + N\sigma_{pn}^{DY} \;\ .
\label{DYpA}
\end{equation}
It has been shown \cite{ABC90,SSG94} that the ratio of the cross
section for the scattering of protons from the nucleus with $N-Z
\neq 0$ to that from an isoscalar target such as deuterium is
sensitive to the $\overline{d}_p(x) - \overline{u}_p(x)$ difference.
These ratios have been measured by the E772 Collaboration at
FNAL\cite{ABC90} for carbon, calcium, iron and tungsten targets.
Neglecting nuclear effects, elementary algebra leads to the
following result for the ratio
\begin{equation}
R_{DY}=
\frac {2 \sigma^{DY} (p+A)} {A \sigma^{DY} (p+d)} \; = \;
\frac{2Z}{A} \; + \; \frac{N-Z}{A} \;
\frac{2\sigma^{pn}(x_1,x_2)}
 { \sigma^{pp}(x_1,x_2) + \sigma^{pn}(x_1,x_2) },
\label{ratio}
\end{equation}
where $Z$, $N$, $A$ are number of protons, neutrons and the atomic
number, respectively. In the large $x_2$ (target) limit the ratio
takes a very simple form \cite{GMA92}:
\begin{equation}
R_{DY}(x)=1+{N-Z\over A}{\Delta(x)\over \bar u(x)+\bar d(x)},
\end{equation}
showing that the Drell-Yan processes with non-isoscalar targets are
relevant for the issue of the asymmetry.

The experimental ratios are consistent with symmetric quark
distributions \cite{GMA92,ABC90} (see Fig.5), which renders those
data useless for establishing the asymmetry. Moreover, using
asymmetric quarks distribution functions (solid and dashed lines)
has a rather small effect on the ratio. The ratio obtained with the
recent $MSR(A)$ quark distributions \cite{MSR94} almost coincides
with the result of our model. As seen from the figure these ratios
do not provide a test sensitive enough.

An alternative idea to study the asymmetry was proposed more than a
decade ago by Ito et al.\cite{IFJ81}. They have suggested to analyze
the logarithmic derivative of the rapidity distribution
\begin{equation}
S(\sqrt{\tau})
= {d\over dy}\left. ln{d^{2}\sigma\over dM dy}\right|_{y=0} \;\ ,
{\rm\ with\ }\tau=x_1 x_2\;,
\label{rapslope}
\end{equation}
where $y = ln(x_{1}/x_{2})/2$ is the rapidity. This quantity also
possesses the desired property of being independent of the
$K$-factor.  In terms of the quark distributions the slope can be
expressed as
\begin{equation}
S(x)={x^2\over X(x,x)}\left(
 {\partial\over\partial x_1}X(x_1,x)\Bigg|_{x_1=x}
-{\partial\over\partial x_2}X(x,x_2)\Bigg|_{x_2=x}
\right)
\;\  ,
\label{rapslopenuc}
\end{equation}
where
\begin{eqnarray}
X(x_1,x_2) &=& \frac{4}{9} \; \lbrace
u(x_1) \bigl[ \frac{Z}{A} \overline{u}(x_2)
            + \frac{N}{A} \overline{d}(x_2) \bigr]
+ \overline{u}(x_1) \bigl[\frac{Z}{A} u(x_2)
+ \frac{N}{A} d(x_2) \bigr]
 \rbrace
\\ \nonumber
  &+& \frac{1}{9} \; \lbrace
d(x_1) \bigl[ \frac{Z}{A} \overline{d}(x_2)
            + \frac{N}{A} \overline{u}(x_2) \bigr]
+ \overline{d}(x_1) \bigl[\frac{Z}{A} d(x_2)
+ \frac{N}{A} u(x_2) \bigr]
 \rbrace
\\ \nonumber
  &+& \frac{1}{9} \; \lbrace
s(x_1) \overline s(x_2)+\overline s(x_1) s(x_2) \rbrace
\; .
\end{eqnarray}

We illustrate the effect of the $\overline{u}-\overline{d}$
asymmetry on the slope of the rapidity distribution in Fig.6.  Here
the solid lines are calculated using the asymmetric quark
distribution of the recent $MSR(A)$ quark parameterization
\cite{MSR94} and those obtained from our meson model
\cite{HSS94,HSS93,HSS95Q}. The dashed lines are obtained by using
symmetrized quark distributions given by
Eq.(\ref{symmetrization}). The arrows show the effect of the
symmetrization which decreases the slope. In this and all following
calculations we have included corrections due to the $Q^2$ scale
changing ($Q^2=sx_1x_2$) by employing leading log Altarelli-Parisi
QCD evolution.
It has turned out that the resulting effects of the evolution are
rather small.

The rapidity slope (see Eq. \ref{rapslope}) is a quantity which is
sensitive not only to the $\overline{u}-\overline{d}$ asymmetry but
also to valence quark distributions. In Fig.7 we display the slope
of the rapidity distribution calculated with different quark
distributions.  The solid line is the result of our meson model
\cite{HSS95Q}. The dotted line is the result obtained with the Owens
parameterization \cite{O91} of the quark distributions, the dashed
line was obtained with the recent MRS(A) parameterization
\cite{MSR94} with $\overline{u}-\overline{d}$ asymmetry and the
dash-dotted line was obtained with MRS($S_{0}^{'}$) \cite{MSR93}
(symmetric) distribution.  Fig.7 clearly demonstrates that the
asymmetry is not the only ingredient and a reasonable description of
the experimental data \cite{IFJ81} can be obtained with both flavour
symmetric and asymmetric distributions.

In obtaining both Eq.(\ref{ratio}) and Eq.(\ref{rapslopenuc}) we
have neglected completely all nuclear effects like Fermi-motion,
nuclear binding, excess pions or shadowing. Although they are
predominantly flavour symmetric, it is obvious that they can modify
the conclusions drawn based on the nuclear data. An information on
nuclear effects can result only from the comparison of the slopes
for a nuclear target and for a deuteron one. It seems essential in
the future to analyze more elementary processes, i.e. dilepton
production in the proton-proton and proton-deuteron collisions.  In
the following we shall concentrate on those reactions as most
reliable source of the information on the flavour asymmetry of the
sea quarks.

A quantity which can be extracted almost directly from the
experiment is
\begin{equation}
A_{DY}(x_1,x_2) =
 \frac{ \sigma_{pp}(x_1,x_2) - \sigma_{pn}(x_1,x_2) }
      { \sigma_{pp}(x_1,x_2) + \sigma_{pn}(x_1,x_2) } \;\ ,
\label{ADY}
\end{equation}
which we will call Drell-Yan asymmetry. In formula (\ref{ADY})
$\sigma_{pp}$ and $\sigma_{pn}$ are the cross sections for the
dilepton production in the proton-proton and proton-neutron
scattering.
The Drell-Yan asymmetry (\ref{ADY}) can be expressed in terms of
$\bar q$ and $\Delta$ introduced in Eq.(\ref{asymdecomp})
\begin{equation}
A_{DY}(x_1,x_2) =
\frac{
[u(x_2) - d(x_2)] [3 \bar q(x_1) - 5/2 \Delta(x_1)]
- [4 u(x_1) - d(x_1)] \Delta(x_2)
}{
[u(x_2) + d(x_2)] [5 \bar q(x_1) - 3/2 \Delta(x_1)]
+ [4 u(x_1) + d(x_1)] 2 \bar q(x_2)
} \; .
\end{equation}
In the case of flavour symmetric sea ($\Delta =0$) it is natural to
expect that $A_{DY} > 0$ since $u > d$.
The sign of $A_{DY}$ can be, however, reversed by increasing the
flavour asymmetry of the proton sea ($\Delta > 0$).

Two-dimensional maps of the Drell-Yan asymmetry as a function of
$x_1$ and $x_2$ are shown in the form of the contour plots in Fig.8.
The different maps were obtained with the Owens parameterization
\cite{O91} (left-upper corner), symmetric MRS($S_{0}^{'}$)
\cite{MSR93} (right-upper corner), the new MRS(A) \cite{MSR94} with
the $\overline{u}-\overline{d}$ asymmetry built in (left-lower
corner) and the prediction of the meson cloud model
\cite{HSS94,HSS93,HSS95Q} (right-lower corner).  The result obtained
with the Owens (symmetric) parameterization and symmetric
MRS($S_{0}^{'}$) parameterization are quite similar.  This clearly
demonstrates that the asymmetry $A_{DY}$ is the desired quantity --
insensitive to the valence quark distributions.  It is also worth
noting here that $A_{DY}$ is positive in the whole range of $(x_{1},
x_{2})$.  How the $\overline{u}-\overline{d}$ asymmetry influences
$A_{DY}$ is shown in two lower panels. It is very promising that
$A_{DY}$ obtained with the asymmetric quark distributions (lower
panels) differs considerably (please note the change of sign in the
lower panels) from the result obtained with symmetric distribution
(upper panels) and this should make an unambiguous verification of
the flavour asymmetry of the sea quarks possible.  It is not random
in our opinion that the result obtained within the meson cloud model
is very similar to that obtained from the parameterization fitted to
different experimental data. We stress in this context that $A_{DY}$
calculated in the meson cloud model is fairly insensitive to the
quark distributions in the bare nucleons (baryons). It is primarily
sensitive to the $\overline{u}-\overline{d}$ asymmetry which is
fully determined by the quark distributions in pions (mesons), taken
here from the pion-nucleus Drell-Yan processes. We have assumed that
the quark distributions in other mesons are related to those for the
pion via $SU(3)_{f}$ symmetry.

Following the suggestion of Ellis and Stirling, the NA51
Collaboration at CERN has measured recently the $A_{DY}$ asymmetry
along the $x_{1} = x_{2}$ diagonal \cite{B94}. Due to low statistics
only $A_{DY}$ at low $x=x_{1}=x_{2}$ was obtained. In Fig.9 we show
their experimental result (one experimental point) together with the
results obtained with different quark distributions.  The meaning of
the lines here is the same as in Fig.6.  The result denoted as MCM,
obtained within the meson cloud model \cite{HSS93,HSS94,HSS95Q}
essentially without free parameters, nicely agrees with the
experimental data point.  In order to better understand the result
and the relation to the $\overline{u}-\overline{d}$ asymmetry let us
express the cross sections in Eq.(\ref{ADY}) in terms of the quark
distributions.  Assuming proton-neutron isospin symmetry and taking
$x_{1}=x_{2}=x$ as for the NA51 experiment one gets in terms of the
quark distributions in the proton
\begin{equation}
A_{DY} =
\frac{ 5(u-d)(\overline{u} - \overline{d}) +
3(u\overline{u} - d\overline{d}) }
{  5(u+d)(\overline{u} + \overline{d}) +
3(u\overline{u} - d\overline{d}) +
4(s\overline{s} + 4c\overline{c}) }.
\label{ADY_qu}
\end{equation}
Let us consider first the case $\bar u=\bar d$. For a crude
estimation one may neglect sea-sea terms (important at small $x$
only) and assume $u_{val}(x)=2d_{val}(x)$, which leads to
$A_{DY}=1/11>0$. The same crude estimate in the case of asymmetric
sea in conjuction with decomposition Eq.(\ref{asymdecomp}) yields
\begin{equation}
A_{DY}={-19\Delta+6\bar q\over -9\Delta+66\bar q}.
\label{simple_ady}
\end{equation}
This demonstates a strong sensitivity both on $\bar d-\bar u$
asymmetry and on the absolute normalization of the sea. The lack of
dependence on the valence quark distributions in the approximate
Eq.(\ref{simple_ady}) suggests a weak dependence in the exact
formula Eq.(\ref{ADY_qu}).  The negative value obtained by NA51
experiment $A_{DY} = -0.09 \pm 0.02 \pm 0.025$ automatically implies
$\overline{d} > \overline{u}$ at least for the measured $x=0.18$
(provided that the proton-neutron isospin symmetry violation is
small(!)).  The data point of the NA51 group is up to now the most
direct evidence for the flavour asymmetry of the sea quarks, which
is explicitly shown in Fig.10 where $A_{DY}$ has been translated into
the ratio of $\overline{u}(x)/\overline{d}(x)$. The $x$ dependence
of the asymmetry is awaiting further experiments. It is expected
that the new experiment planned at Fermilab \cite{G92} will be very
useful in this respect and will provide the $x$ dependence of the
$\overline{u}-\overline{d}$ asymmetry up to $x=0.4$ and will shed
new light on the microscopic structure of the nucleon.  The
meson-cloud model gives definite predictions for the asymmetry
awaiting future experimental verification.

\pagebreak
\section{Conclusions}
\label{Conclusions}

\mbox{ }\indent The violation of the Gottfried Sum Rule observed by
NMC \cite{A91,A'94} together with negative Drell-Yan asymmetry
measured recently \cite{B94} by the NA51 group at CERN give a
support to the conclusion that the $SU(2)$ symmetry of the nucleon
sea is violated.  As discussed recently by Forte \cite{F93} there
are two possible kinds of symmetry violations, called $SU(2)_{Q}$
and $SU(2)_{I}$.  The first one is simply connected with the
asymmetry of light sea antiquarks $\overline{u} - \overline{d}$ in
the proton. The second is related to the violation of the
proton-neutron isospin symmetry. There are no a priori reasons,
except of customs of practitioners in deep-inelastic scattering, for
either symmetry to be more fundamental.  Both GSR violation and
negative Drell-Yan asymmetry can in principle be explained by either
$SU(2)_Q$ ($\overline{d} > \overline{u}$) or/and $SU(2)_I$ (more
abundant neutron than proton sea) symmetry violation.  Some
theoretical arguments \cite{F93} suggest, however, that the
violation of the $SU(2)_{Q}$ symmetry seems to be more probable.
While at present models explaining the $\overline{u}-\overline{d}$
asymmetry have been constructed, no reliable models explaining the
proton-neutron isospin symmetry violation exist.  The proton-neutron
symmetry violation can be expected on the basis of the bag models as
due to the mass difference of the $u$ and $d$ quarks as well as the
corresponding di-quark states. At present a reasonable results can
only be obtained for the valence quarks \cite{RTL93}, which,
however, has no influence on the GSR violation and rather little
effect for the Drell-Yan asymmetry at the experimentally measured $x
\approx 0.2$.

The old concept of the meson cloud in the nucleon gives a natural
explanation of the $\overline{u}-\overline{d}$ asymmetry.  The
essential parameters of the model -- coupling constants -- are well
known from the low-energy physics. If the remaining parameters of
the model (cut-off parameters of the vertex form factors) are fixed
from the high-energy neutron and $\Delta^{++}$ production
\cite{HSS94} then both the GSR violation and the Drell-Yan asymmetry
can be well described. The same model gives also a good description
of the neutron electric form factor \cite{HSS94EM}.  In our model
the virtual mesons influence the static properties of the nucleon
(axial vector coupling constant \cite{HSS94}, electromagnetic radii
\cite{HSS94EM}, etc.).

Our model has to be contrasted to the solution of Ball and
Forte\cite{BF93,BBFG94} where mesons are produced radiatively via
modified Altarelli-Parisi equations. Therefore their approach
predicts strong dependence of the Gottfried Sum Rule on the scale,
in the range of intermediate $Q^2$.  In our approach the Gottfried
Sum Rule is constant, at least in the leading logarithm
approximation. It would be very important to test these two
scenarios experimentally. A preliminary evaluation of the NMC data
seems to support rather our model \cite{Br95}.

We have discussed possibilities to identify the $\overline{u}-
\overline{d}$ asymmetry through the observation of the dilepton
pairs in the hadronic collisions. The analysis of the present
$\mu^+\mu^-$ pair creation data in proton-nucleus scattering is not
conclusive. The rapidity slope, very sensitive to the flavour
asymmetry, depends also on the valence quark distributions. The
ratio of the cross section in proton-nucleus to that in
proton-deuteron collision is compatible with symmetric sea quark
distributions. However, the case of asymmetry concentrated at rather
small $x$ is not excluded. The meson cloud model gives results
compatible with the E772 Fermilab experimental data
\cite{SSG94}. Elementary nucleon-nucleon Drell-Yan processes seem to
be much better suited to study the $\bar d-\bar u$ asymmetry.

The presence of virtual mesons in the nucleon, especially pions, can
explain the new result of the NA51 group at CERN for the Drell-Yan
asymmetry. The Drell-Yan asymmetry is a quantity fairly insensitive
to the valence quark distribution and very sensitive to the flavour
asymmetry of the sea. The Drell-Yan asymmetries obtained with
different valence quark distributions and symmetric sea are similar
and positive. The meson cloud model predicts negative $A_{DY}$,
which is consistent with the only existing experimental data point
\cite{B94}. Furthermore it gives definite prediction for the $x$
dependence of the flavour asymmetry, awaiting experimental
verification.  The new experiment planned at Fermilab \cite{G92}
will open such a possibility.

\vskip 2cm

One of us (A.S.) is very grateful to Peter Sutton for a very useful
discussion on higher-order QCD corrections. M.E. thanks the
Humboldt Foundation for her stay in Germany.



\begin{thebibliography}{10}

\bibitem{G67}
K. Gottfried.
\newblock {\em Phys. Rev. Lett.\ }{\bf 18} (1967) 1174.

\bibitem{A91}
P. Amaudruz {\it et al}.
\newblock {\em Phys. Rev. Lett.\ }{\bf 66} (1991) 2712.

\bibitem{A'94}
M. Arneodo {\it et al}.
\newblock {\em Phys. Rev.\ }{\bf D50} (1994) R1.

\bibitem{BK92}
B. Bade{\l}ek and J. Kwieci\'nski.
\newblock {\em Nucl. Phys.\ }{\bf B370} (1992) 278.

\bibitem{Z92A}
V.R. Zoller.
\newblock {\em Z. Phys.\ }{\bf C54} (1992) 425.

\bibitem{KH93}
H. Khan and P. Hoodhbhoy.
\newblock {\em Phys. Lett.\ }{\bf B298} (1993) 181.

\bibitem{MT93}
W. Melnitchouk and A.W. Thomas.
\newblock {\em Phys. Rev.\ }{\bf D47} (1993) 3794.

\bibitem{KU91}
L.P. Kaptari and A.Yu. Umnikov.
\newblock {\em Phys. Lett.\ }{\bf B272} (1991) 359.

\bibitem{B94}
{A. Baldit {\it et al}}.
\newblock {\em Phys. Lett.\ }{\bf B332} (1994) 244.

\bibitem{SSG94}
A. Szczurek, J. Speth and G.T. Garvey.
\newblock {\em Nucl. Phys.\ }{\bf A570} (1994) 765.

\bibitem{HNSS94}
H. Holtmann, N.N. Nikolaev, J. Speth and A. Szczurek.
\newblock {\em {J\"ulich preprint KFA-IKP(TH)-14}\ }
(1994) submitted to {\it Z. Phys. A.}

\bibitem{IFJ81}
A.S.Ito {\it et al}.
\newblock {\em Phys. Rev.\ }{\bf D23} (1981) 604.

\bibitem{ET84}
M. Ericson and A.W. Thomas.
\newblock {\em Phys. Lett.\ }{\bf B148} (1984) 181.

\bibitem{G92}
G.T. Garvey {\it et al}.
\newblock {\em FNAL proposal, P866\ } (1992) .

\bibitem{A92}
P. Amaudruz {\it et al}.
\newblock {\em Phys. Lett.\ }{\bf B292} (1992) 159.

\bibitem{F93}
S. Forte.
\newblock {\em Phys. Rev.\ }{\bf D47} (1993) 1842.

\bibitem{MSG93}
{Bo-Qiang Ma, A. Sch\"afer, W. Greiner}.
\newblock {\em Phys. Rev.\ }{\bf D47} (1993) 51.

\bibitem{AP77}
G. Altarelli and G. Parisi.
\newblock {\em Nucl. Phys.\ }{\bf B126} (1977) 298.

\bibitem{RS79}
D.A. Ross and C.T.Sachrajda.
\newblock {\em Nucl. Phys.\ }{\bf B149} (1979) 497.

\bibitem{HM90}
E.M. Henley and G.A. Miller.
\newblock {\em Phys. Lett.\ }{\bf B251} (1990) 453.

\bibitem{SST91}
A. Signal, A.W. Schreiber and A.W. Thomas.
\newblock {\em Mod. Phys. Lett.\ }{\bf A6} (1991) 271.

\bibitem{MTS91}
W. Melnitchouk, A.W. Thomas and A.I. Signal.
\newblock {\em Z. Phys.\ }{\bf A340} (1991) 85.

\bibitem{K191}
S. Kumano.
\newblock {\em Phys. Rev.\ }{\bf D43} (1991) 59.

\bibitem{HSB91}
W-Y.P. Hwang, J. Speth and G.E. Brown.
\newblock {\em Z. Phys.\ }{\bf A339} (1991) 383.

\bibitem{Z92}
V.R. Zoller.
\newblock {\em Z. Phys.\ }{\bf C53} (1992) 443.

\bibitem{SS93}
A. Szczurek and J. Speth.
\newblock {\em Nucl. Phys.\ }{\bf A555} (1993) 249.

\bibitem{HSS94}
H. Holtmann, A. Szczurek and J. Speth.
\newblock {\em {J\"ulich preprint KFA-IKP(TH)-25}\ }
(1994) submitted to {\it
  Phys. Rev. }{\bf D}.

\bibitem{ES91}
S.D. Ellis and W.J. Stirling.
\newblock {\em Phys. Lett.\ }{\bf B256} (1991) 258.

\bibitem{GMA92}
P.L. McGaughey {\it et al}.
\newblock {\em Phys. Rev. Lett.\ }{\bf 69} (1992) 1726.

\bibitem{MSR94}
A.D. Martin, W.J. Stirling and R.G. Roberts.
\newblock {\em Phys. Rev.\ }{\bf D50} (1994) 6734.

\bibitem{HSS93}
H. Holtmann, A. Szczurek and J. Speth.
\newblock {\em {J\"ulich preprint KFA-IKP(TH)-1993-33},\ }

\bibitem{CL93}
{T.D. Cohen and D.B. Leinweber}.
\newblock {\em Comments Nucl. Part. Phys.\ }{\bf 21} (1993) 137.

\bibitem{S72}
J.D. Sullivan.
\newblock {\em Phys. Rev.\ }{\bf D5} (1972) 1732.

\bibitem{SMRS92}
P.J. Sutton, A.D. Martin, R.G. Roberts and W.J. Stirling.
\newblock {\em Phys. Rev.\ }{\bf D45} (1992) 2349.

\bibitem{HS92}
W-Y.P. Hwang and J. Speth.
\newblock {\em Phys. Rev.\ }{\bf D46} (1992) 1198.

\bibitem{HSS95Q}
H. Holtmann, A. Szczurek and J. Speth.
\newblock {\em paper in preparation\ }.

\bibitem{Le93}
W.C. Leung.
\newblock {\em Phys. Lett.\ }{\bf B317} (1993) 655.

\bibitem{M92}
S.R. Mishra {\it et al}.
\newblock {\em Phys. Rev. Lett.\ }{\bf 24} (1992) 3499.

\bibitem{DY71}
S.D. Drell and T.M. Yan.
\newblock {\em Ann. Phys.\ }{\bf 66} (1971) 578.

\bibitem{O91}
J.F. Owens.
\newblock {\em Phys. Lett.\ }{\bf B266} (1991) 126.

\bibitem{MSR93}
A.D. Martin, W.J. Stirling and R.G. Roberts.
\newblock {\em Phys. Lett.\ }{\bf B306} (1993) 145.

\bibitem{ABC90}
D.M. Alde {\it et al}.
\newblock {\em Phys. Rev. Lett.\ }{\bf 64} (1990) 2479.

\bibitem{RTL93}
E.N. Rodionov, A.W. Thomas and J.T. Londergan.
\newblock {\em Mod. Phys. Lett.\ }A9 (1994) 1799.

\bibitem{HSS94EM}
H. Holtmann, A. Szczurek and J. Speth.
\newblock {\em {J\"ulich preprint KFA-IKP(TH)-1994-33},\ }
submitted to {\it Z.  Phys. }{\bf A}.

\bibitem{BF93}
R.D. Ball and S. Forte.
\newblock {\em Nucl. Phys.\ }{\bf B425} (1994) 516.

\bibitem{BBFG94}
{R.D. Ball, V. Barone, S. Forte, M. Genovese}.
\newblock {\em Phys. Lett.\ }{\bf B329} (1994) 505.

\bibitem{Br95}
{A. Br\"ull}.
\newblock {Habilitationsschrift, Heidelberg 1995}.

\end{thebibliography}

\begin{figure}
\caption{ The $x$-dependence of $F_{2}^{ep}(x) - F_{2}^{en}(x)$. The
solid line was obtained including the effect of virtual mesons. The
data points at $Q^2$ = 4 (GeV/c)$^2$ are taken from
Ref.\protect\cite{A'94}.}
\end{figure}

\begin{figure}
\caption{The total sea quark distribution
$x (\bar u(x) + \bar d(x) + \bar s(x))$ obtained from our model. The
solid line is the result of the full calculation, i.e. including
the sea quarks in the bare nucleons (see Eq. \protect\ref{baredis}). For
comparison by the dashed line we show result obtained when neglecting
the sea quark distribution in the bare baryons (mesonic contribution).
The experimental data was taken from \protect\cite{M92}.}
\end{figure}

\begin{figure}
\caption{ The $K$-factors for the dilepton production in $pp$ (solid
line) and $pn$ (dashed line) collisions for two different leading
order quark distributions taken from \protect\cite{O91} and
\protect\cite{MSR93}. }
\end{figure}

\begin{figure}
\caption{ The cross section $M^{3} {d^2 \sigma}/{dx_{F} dM}$ for the
production of the dilepton pairs in the proton-deuteron collisions.
Shown is the fit of the $K$-factors for various quark distributions
to the experimental data \protect\cite{GMA92}.}
\end{figure}

\begin{figure}
\caption{ The Drell-Yan ratio defined by Eq.(\protect\ref{dyratio})
for iron and tungsten. The solid line is the result of our model,
the dashed line the $MSR(A)$ parametrization and the dash-dotted
line the $MSR(S'_0)$ parameterization.}
\end{figure}

\begin{figure}
\caption{ The effect of the $\overline{u}-\overline{d}$ asymmetry on
the slope of the rapidity distribution. The two solid lines were
calculated with the asymmetric quark distribution of the recent
$MRS(A)$ quark parameterization \protect\cite{MSR94} and that
obtained from our model (MCM) \protect\cite{HSS93,HSS94,HSS95Q}.
The dashed lines were obtained using the symmetrization procedure
(see Eq. \protect\ref{symmetrization}). The arrows show the effect
of the symmetrization in both cases.}
\end{figure}

\begin{figure}
\caption{ The slope of the rapidity distribution. The solid line is
the result of our model \protect\cite{HSS95Q}. The dotted line is
the result obtained with the Owens parameterization
\protect\cite{O91}, the dashed line was obtained with the $MRS(A)$
parameterization \protect\cite{MSR94} with $\overline{d} -
\overline{u}$ asymmetry and the dash-dotted line was obtained using
the $MRS(S_{0}^{'})$ (symmetric) distribution.}
\end{figure}

\begin{figure}
\caption{ A two-dimensional map of the Drell-Yan asymmetry as a
function of $x_1$ and $x_2$. Shown are results obtained with the
Owens parameterization \protect\cite{O91} (left-upper corner),
symmetric $MSR(S_{0}^{'})$ (right-upper corner), new $MRS(A)$ with
the $\overline{u} - \overline{d}$ asymmetry built in (left-lower
corner) and prediction of the meson cloud model
\protect\cite{HSS93,HSS94,HSS95Q} (right-lower corner). Note the
change of the sign in the lower panels. }
\end{figure}

\begin{figure}
\caption{ The Drell-Yan asymmetry along the $x_{1} = x_{2}$
diagonal.  The meaning of the lines here is the same as in Fig.6.
The experimental data point is taken from the recent result of the
NA51 Collaboration at CERN \protect\cite{B94}.}
\end{figure}

\begin{figure}
\caption{ The $\overline{u}(x)/\overline{d}(x)$ ratio as obtained
from the meson cloud model (solid line) \protect\cite{HSS95Q} and
the $MSR(A)$ parameterization (dashed line) compared with the
experimental result of the NA51 collaboration \protect\cite{B94}.}
\end{figure}

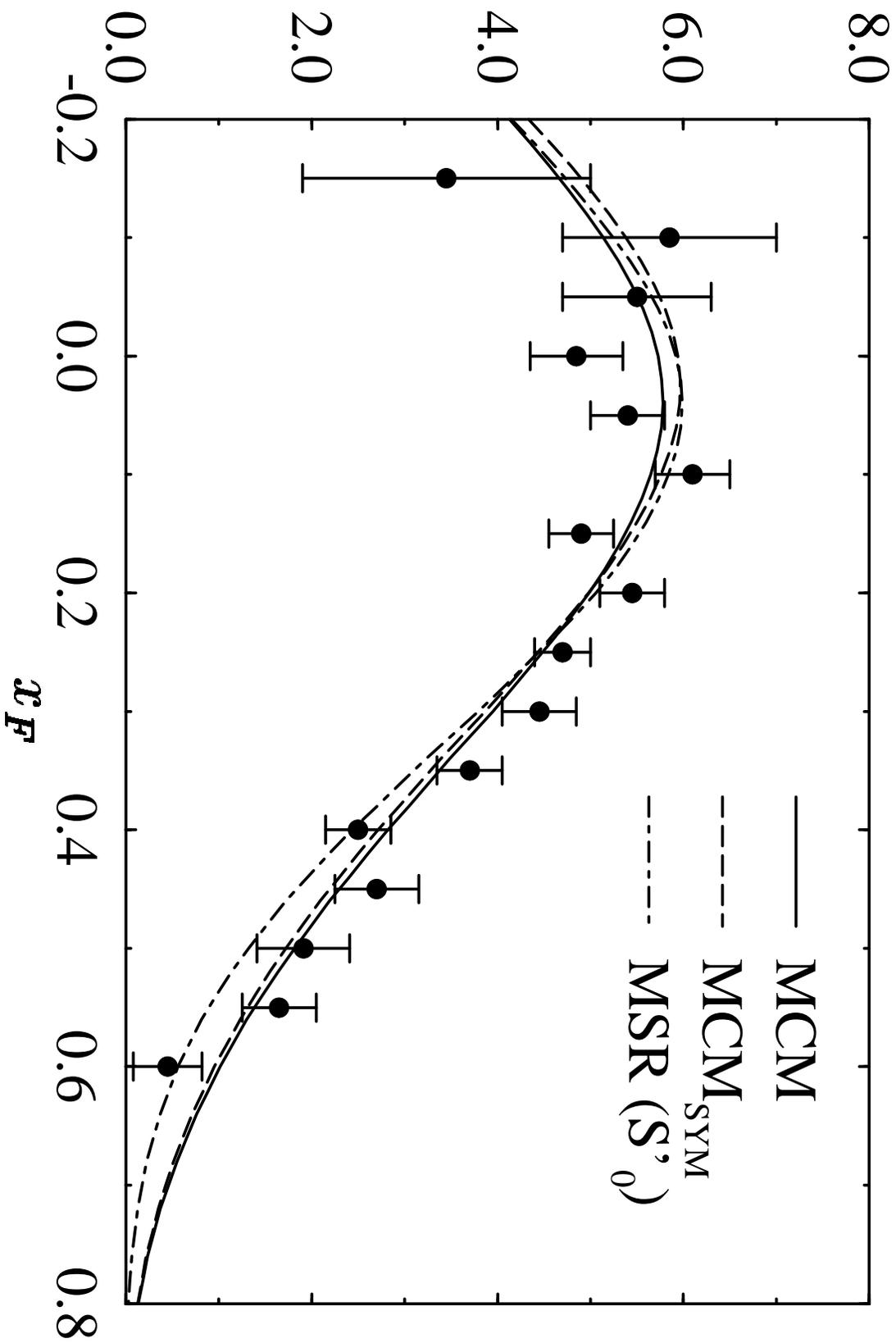
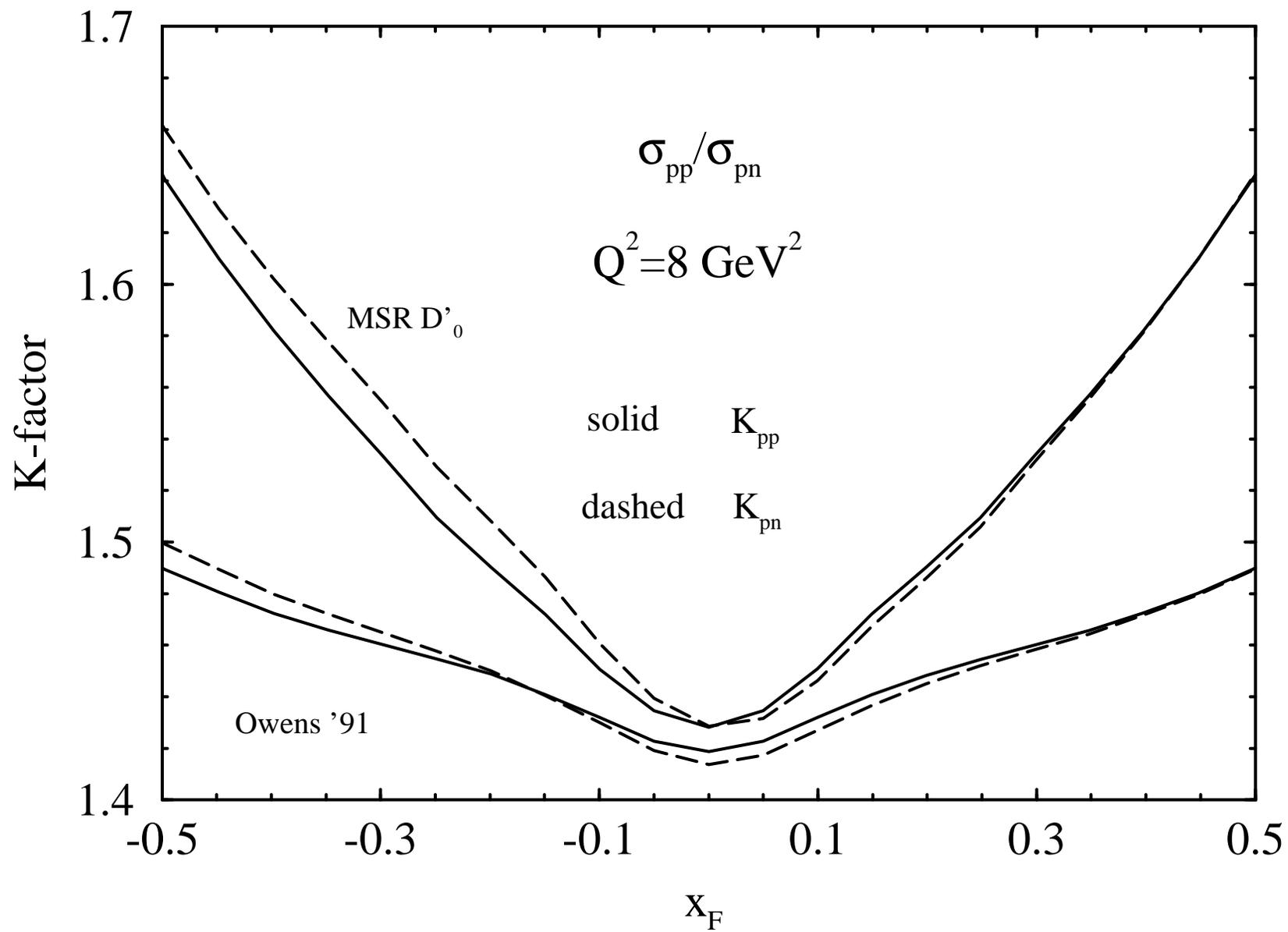
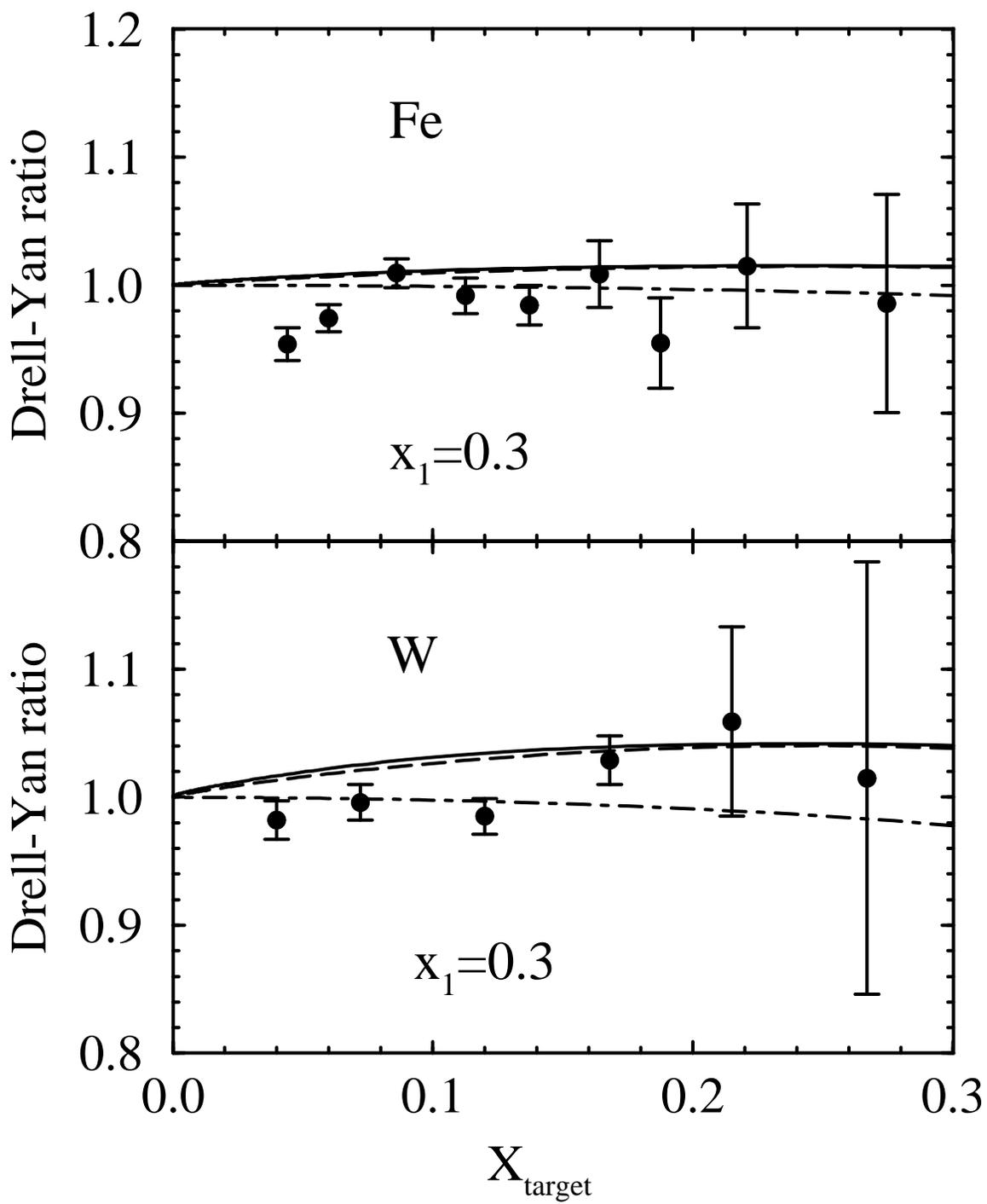
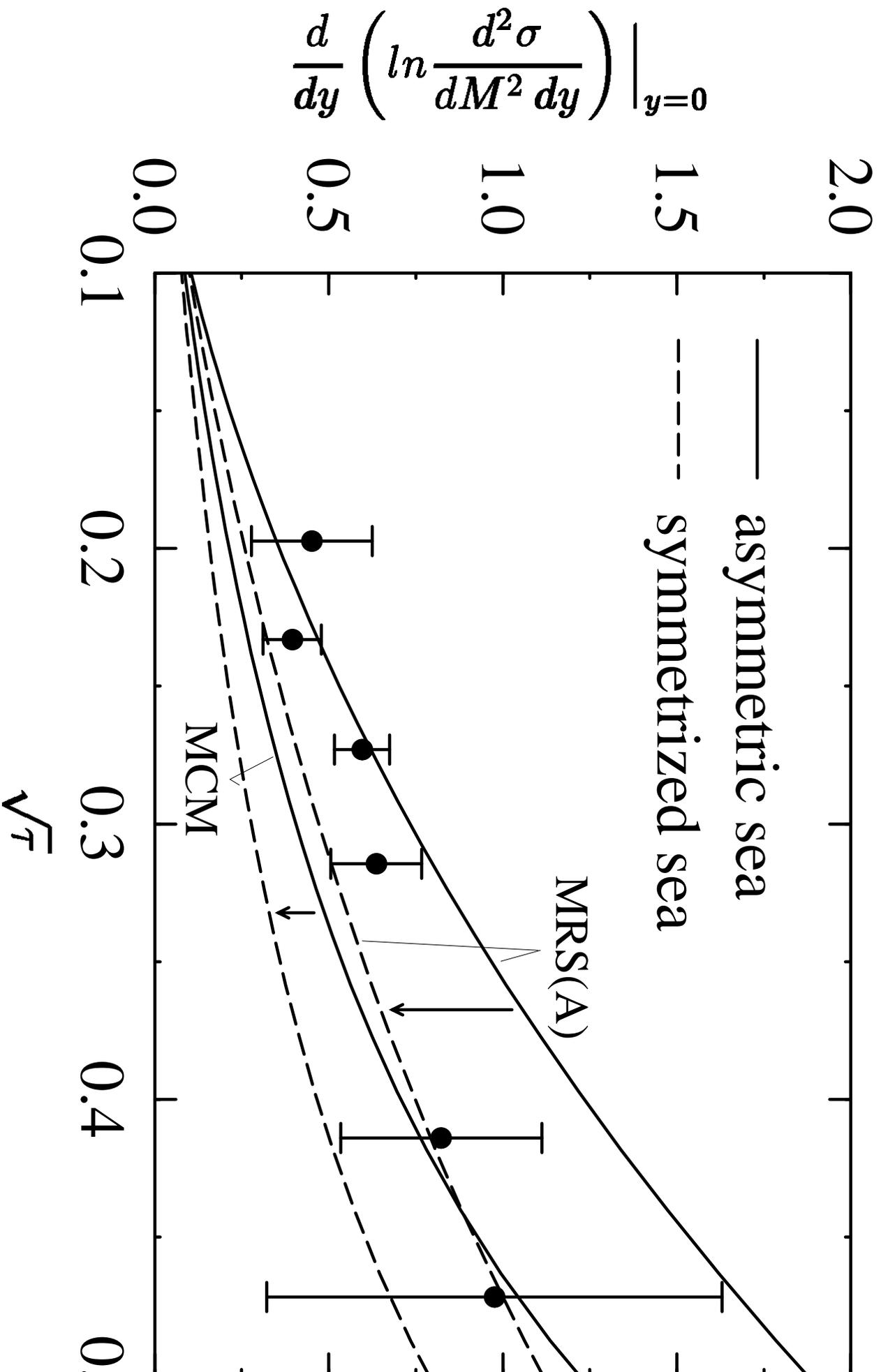
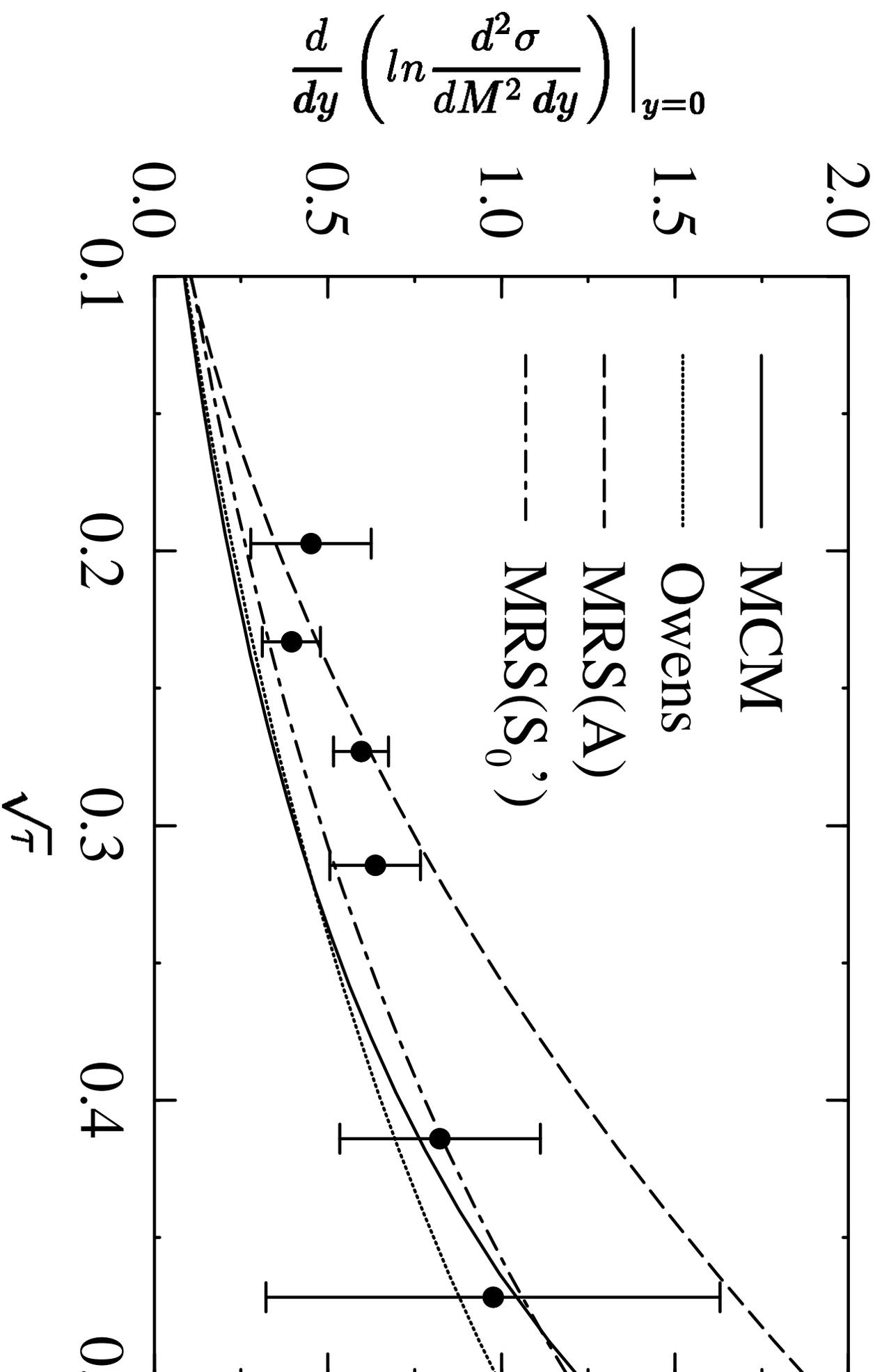
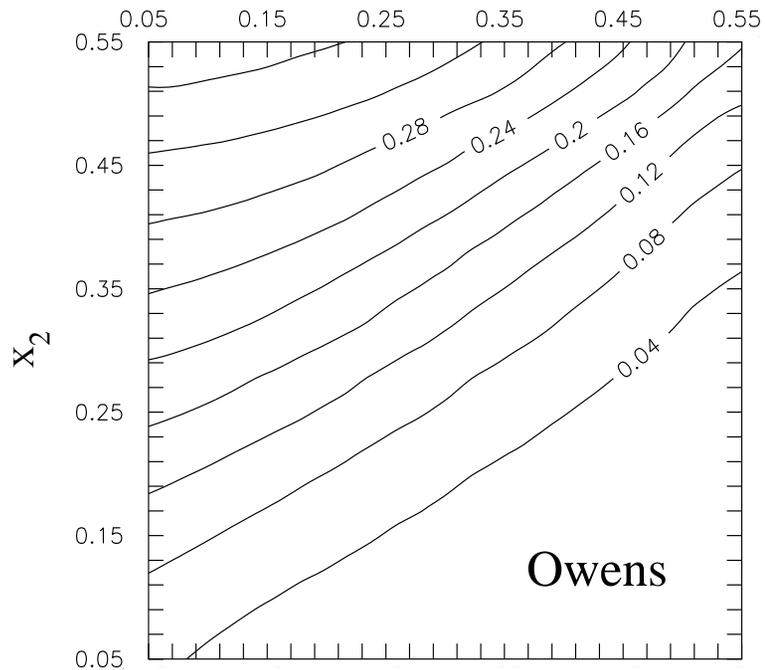
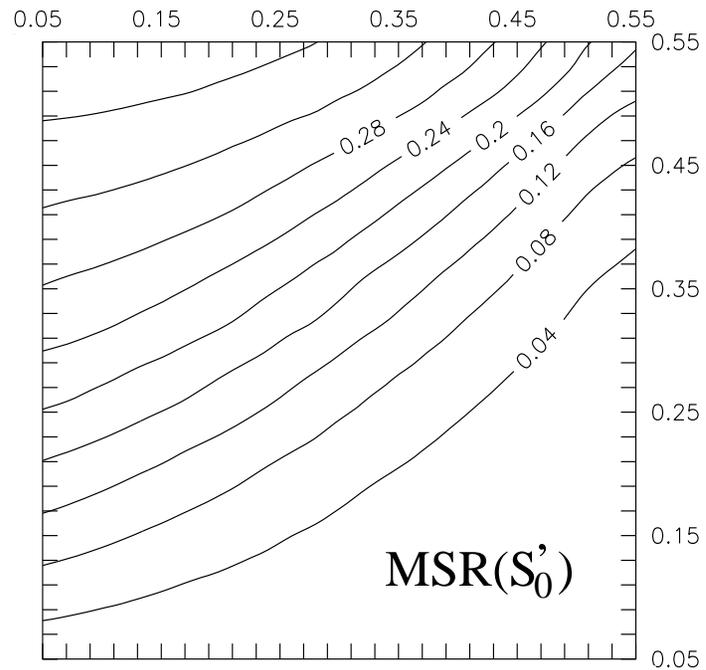
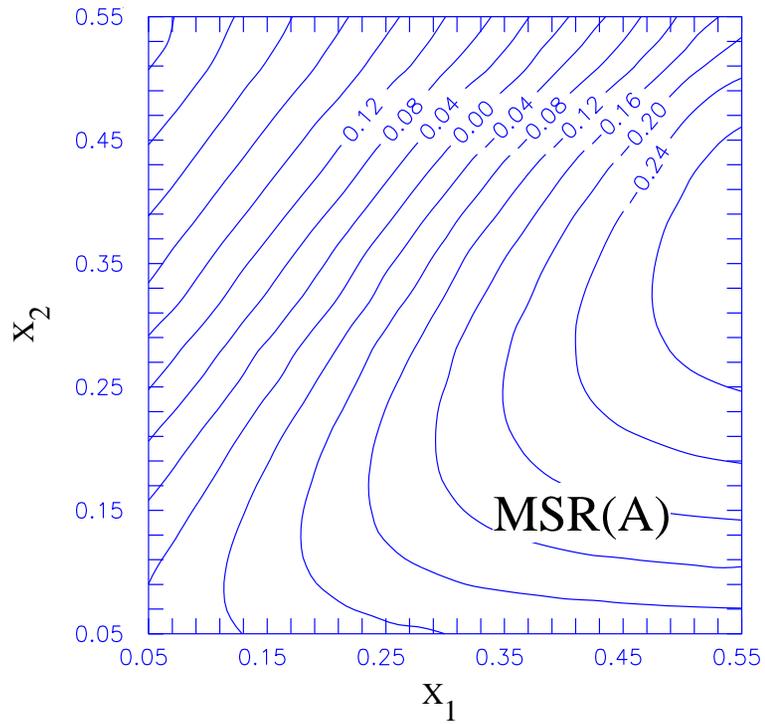
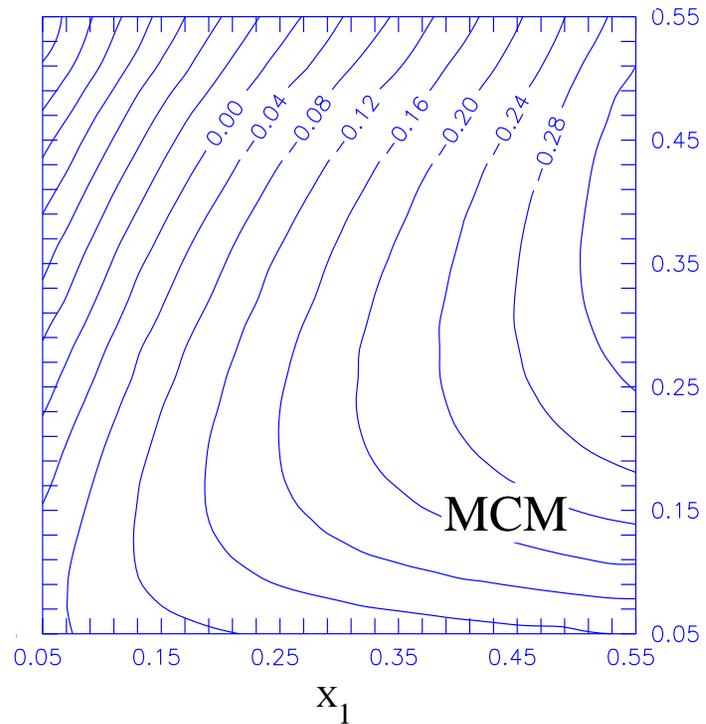
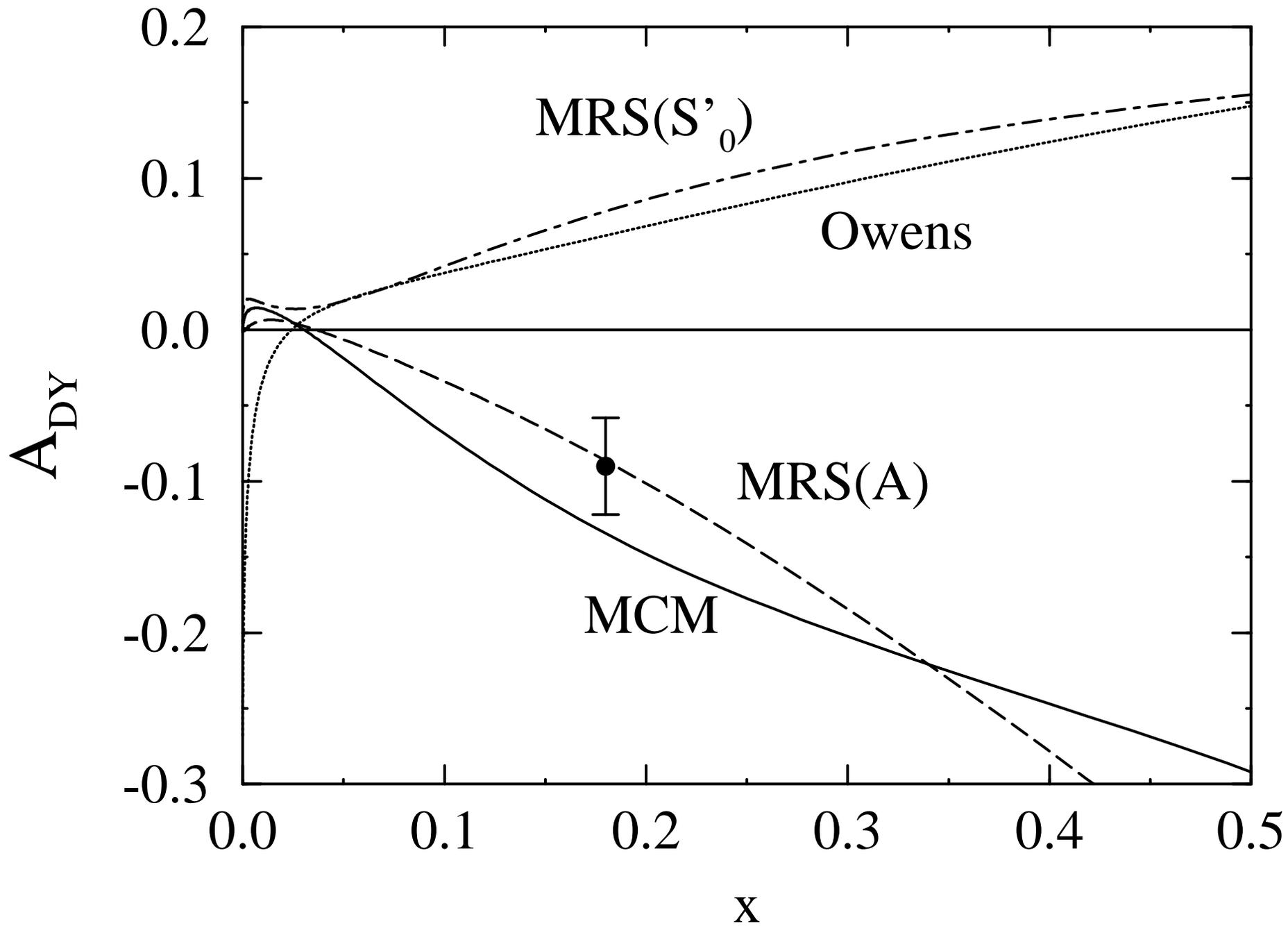
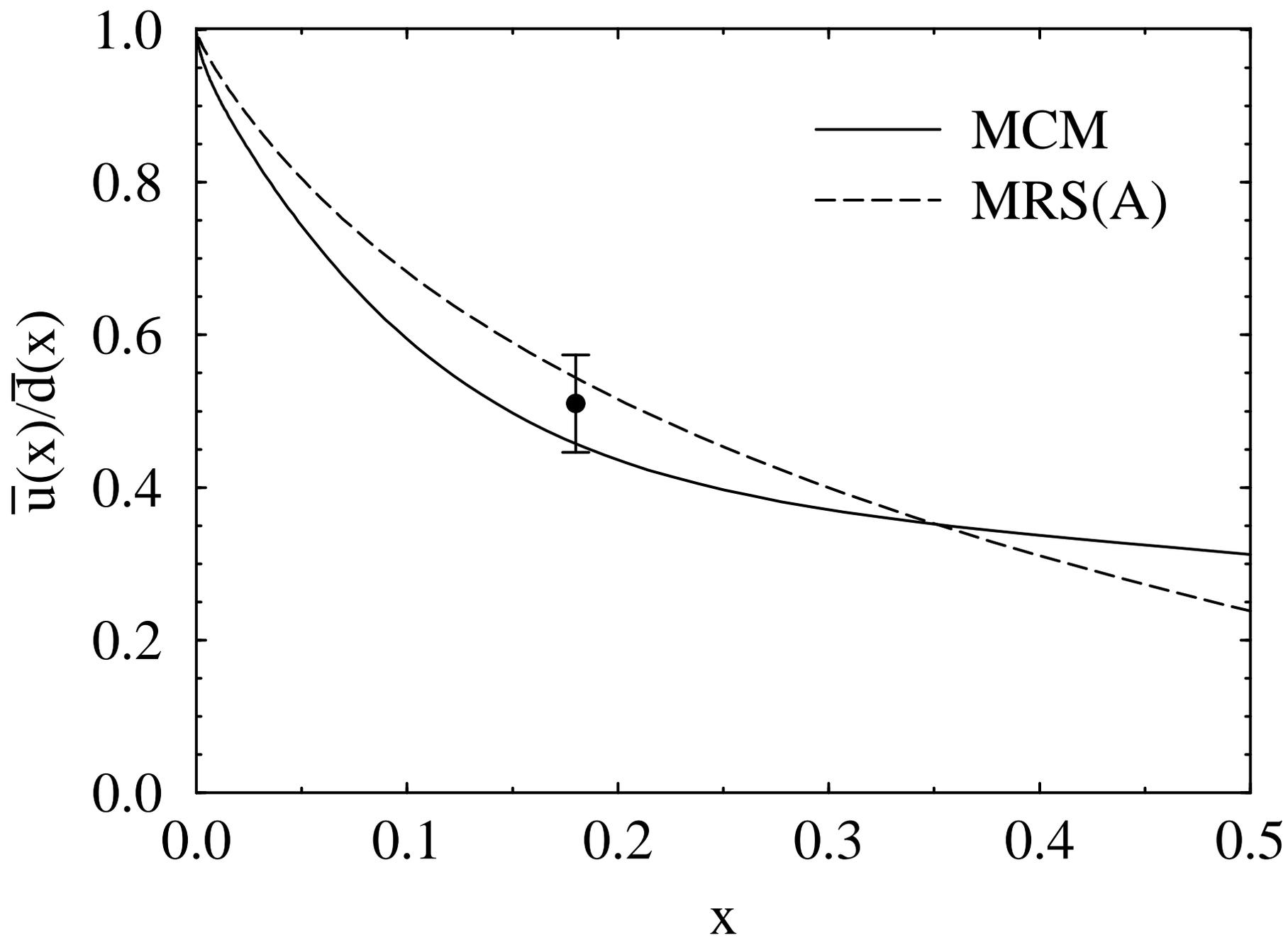
\end{document}